# Evaluation of Query Generators for Entity Search Engines

Stefan Endrullis      Andreas Thor      Erhard Rahm

University of Leipzig
Germany

{endrullis, thor, rahm}@informatik.uni-leipzig.de

## ABSTRACT
Dynamic web applications such as mashups need efficient access to web data that is only accessible via entity search engines (e.g. product or publication search engines). However, most current mashup systems and applications only support simple keyword searches for retrieving data from search engines. We propose the use of more powerful search strategies building on so-called query generators. For a given set of entities query generators are able to automatically determine a set of search queries to retrieve these entities from an entity search engine. We demonstrate the usefulness of query generators for on-demand web data integration and evaluate the effectiveness and efficiency of query generators for a challenging real-world integration scenario.

## 1. INTRODUCTION
There is a huge number of deep web sources whose content is hidden behind (entity) search engine interfaces [Be01] and, thus, can only be accessed via suitable search queries. Entity search engines (ESE) are a popular way to access this valuable data, e.g., product search engines (Google Product Search, Yahoo Shopping) or bibliographic search engines for scientific publications (Google Scholar, Microsoft Libra). ESEs are not only employed by human users but are increasingly used by computer programs such as mashup applications. Mashups combine content from multiple (web) sources and services in a dynamic fashion, i.e., data integration occurs at runtime (on demand) based on specific user input. Web data access builds on existing (large) deep web sources that are accessible via web APIs and on information extraction methods, e.g., based on screen scraping.

One challenge in using ESEs for data integration is that these data sources need to be accessed via specific search query interfaces. Current mashup systems and applications are typically limited to simple keyword searches for retrieving data from search engines, making it difficult to obtain good results with high recall and precision. We argue that better query results with acceptable performance require the exploitation of specific search functionality of entity search engines. Furthermore, the data quality of large-scale ESEs may be limited requiring a post-processing effort, e.g., to identify relevant result entities or deal with duplicate results.

We focus on common integration scenarios where a specific set of entities needs to be found in an ESE, e.g., to obtain more information on the entities for further processing. For example, in the e-commerce domain we may have a list of products to be searched for to identify suppliers with the lowest price per product or to obtain corresponding product reviews. In the bibliographic domain, we may want to obtain citation data for a given list of papers, e.g., to identify the top-cited papers of a conference or of an author.

To solve such integration tasks we want to exploit existing ESEs in an efficient way. For illustration we can consider the fictional publication entities of Table 1 that shall be found at Google Scholar. The naïve approach of using one keyword query per input object generally results in a high number of queries and may still not retrieve all relevant entities. Hence, the challenge is to determine suitable search queries for a given set of entities which achieve both good result quality (in terms of recall and precision w.r.t. the entities of interest) as well as good runtime performance. For example, very specific search queries (e.g., using the exact product name) may miss relevant entries in the presence of name variations whereas relaxed queries may suffer from many irrelevant results. Furthermore, the number of search queries should be minimized to support sufficiently fast response times.

Finding the most effective and efficient set of search queries is a difficult optimization problem depending on many factors, especially the particular set of input entities as well as on the ESEs to be used. As a first step to solve the problem we propose and evaluate the use of different query generators per ESE each of which can automatically generate suitable search queries to find a given set of entities. Query generators may use simple keyword searches but can also utilize specific search features for improving effectiveness or efficiency. In particular, query generators can search for multiple entities simultaneously to reduce the number of queries.

After a brief discussion of related work, we make the following contributions:

- We introduce the concept of query generators to automatically generate search queries for a given set of entities. We provide a generic model for the construction of search queries taking the search capabilities of a search engine into account. Furthermore, we support the generation of queries to simultaneously search for multiple input entities (Section 3).

**Table 1. Fictional example publications**

| Id | Authors | Title |
|---|---|---|
| $s_1$ | {Smith, Jones} | The question to 42 |
| $s_2$ | {Williams, Smith} | Don't Panic! |
| $s_3$ | {Taylor} | The Hitchhiker's Guide to the Galaxy |

- We propose a generic and flexible approach for the evaluation of query generators. Our approach is applicable for different entity search engines and can evaluate multiple query generators in a fully automatic way. We propose the effectiveness measure Coverage and its combination with an efficiency measure to assess the quality of query generators (Section 4).

- We give results of a first evaluation of query generators in the bibliographic domain using the publication search engine Google Scholar. We thereby demonstrate the usefulness of our approach and show how a careful evaluation can be used to identify the most promising query generators for different scenarios (Section 5).

## 2. RELATED WORK

According to [Be01] the deep web is much larger than the surface web making the utilization of deep web sources an important issue in web data management. [Be01] also reports on hundreds of thousands deep websites – along with hundreds of thousands search interfaces.

The automatic generation of search queries for entity search engines has been considered so far from two points of view. First, queries are generated for automatically crawling the hidden web [BF04][RG01]. While our query generators focus on on-demand data access, hidden web crawling usually is an offline process aiming at downloading large portions of the "hidden databases". Second, virtual data integration approaches such as MetaQuerier [CHZ05] have been proposed. These approaches translate queries posed against a global or federated schema into a set of equivalent queries for the underlying hidden web sources. However, these integration approaches require a global schema and an initial user query for query transformation whereas our approach is instance-driven by generating queries based on a given set of entities.

Similar to our approach [TSK07] presents the Karma system that automatically completes a data table that was partially filled by a user beforehand. To that end, appropriate queries are generated based on the given user input. While we focus on ESEs, Karma generates SQL queries for RDBMS and does not consider the specifics of ESEs, e.g., varying data quality and reduced query capabilities in comparison to SQL.

In [JWG06] keyword queries are generated for a set of related source documents. The objective is to find the source documents in a given corpus of documents. While [JWG06] uses these queries to evaluate given document retrieval algorithms, we focus to comparatively assess different query generators themselves to find the best performing ones.

To the best of our knowledge, our work is the first using query generators to determine different types of ESE queries for a given set of entities. In our OCS prototype [TAR07] we already used a fixed set of search queries for one search engine. Here, we propose the general concept of query generators and present an automatic evaluation approach for query generators.

**Table 2. Overview of query capabilities for selected ESEs**

| Capability | Google Scholar | Google Product Search | Amazon (for Books) | Ebay |
|---|---|---|---|---|
| Search predicates $p_i$ | Intitle, Author, Publisher, Year, *free* | Name, Description, Price, *free* | Title, Author, ISBN, Date, Publisher, *free* (+ several categories) | Title, Description, Price, Seller, *free*, (+ several categories) |
| Search values $v_i$ | value, keywords, phrases, pattern | value, keywords, phrases, pattern | value, keywords, phrases | value, keywords, phrases |
| Aggregation(OR) | yes | yes | yes | yes |
| Max. #entities per request | 100 | 100 | 12 | 200 |

## 3. QUERY GENERATORS

We assume the following general **entity search engine model**. A search engine $E$ supports a set of $m$ search predicates $p_1, …, p_m$. Every search predicate typically corresponds to a condition in a search form. For example, the bibliographic search engine Google Scholar supports a general free text predicate as well as specific search predicates (advanced search) for author, title, and publication year. For simplicity, we assume a basic search query $q$ is a conjunction $p_1(v_1) \wedge … \wedge p_m(v_m)$ specifying a matching condition for search values $v_i$ for search predicates $p_i$. Typically only a subset of the available search predicates is used (i.e., a search value $v_i=\varepsilon$ is possible). For search engines, the conjunction of predicates is not necessarily executed as a strict logical AND but the search result may actually contain entities matching only some of the specified predicates. Depending on the search engine capabilities, the search values $v_i$ may represent a single value, a set of keywords, an exact phrase, or a pattern utilizing wildcard symbols. Furthermore the search engine may allow the combination of basic queries with AND or OR. Combining several basic queries is an important feature to reduce the overall number of posed queries and thus to improve the efficiency of search engine access.

Table 2 summarizes the query capabilities for some popular ESEs. They differ in the number and type of predicates as well as in the kind of valid search values. All search engines support a free (unrestricted) search predicate corresponding to a simple search form, e.g., for searching keywords or phrases. The Google ESEs allow the search for (string) patterns by using wildcard symbols whereas Amazon and EBay do not. These search engines support the disjunction of multiple queries but this feature may be subject to some restrictions, e.g., the length of the combined query string. In general, the query capabilities of a search engine may be specified manually but could also be determined automatically [ZHC04, HML+07]. A further important search engine characteristic is the maximal number $z$ of resulting entities per request. It may influence the query generation process since effective query generators try to identify all relevant entities with a minimal number of requests.

Query generators utilize the capabilities of ESEs to generate a certain kind of search queries for a given set of entities. They can implement similar search strategies to the ones used by humans to quickly find certain entities. For example, to find publication [TR07] in Google Scholar one could search for the complete title or search for a combination of author and relevant title keywords, e.g., Thor MOMA. In general a **query generator** takes as input a set $S$ of $n$ entities of the same type (e.g., product, publication, or person). The query generator then generates $k$ queries for a search engine $E$. The goal is that the corresponding query results match the input entities as good as possible, i.e., it aims for a high recall (all relevant entities appear in the result) at a good precision (few irrelevant results). In contrast to manually specified queries, query generators may try to find *multiple entities simultaneously* with one query to reduce the number of queries and thus improve performance. For example, it is more efficient to pose one query returning all relevant results for 10 input entities than to use 10 queries each returning only one or a few relevant results.

We assume that the input entities for query generators are represented as tuples of a relation with a set of attributes $a_1, \ldots, a_u$. Attributes may be single- or multi-valued, e.g., the list of author names for a publication. Each query generator uses the input entities to automatically generate search queries according to four specifications:

- The **partitioning strategy** determines how the input set $S$ is split into subsets $S_1, \ldots, S_k$. One query will be generated for each of the subsets and, thus, $k$ queries are generated for the entire set $S$. Our framework currently supports a naïve and a frequent-value strategy. The *naïve strategy* generates one basic query per entity, i.e. it uses partitions of size 1. This approach is quite expensive but always applicable. In contrast, the *frequent-value strategy* aims at reducing the number of basic queries by identifying search values (attribute values) covering several entities. We use a variation of the well-known Apriori algorithm [AS94] to determine the most frequent attribute values, e.g., publication authors or title keywords, which occur in a minimal number of entities. The entities covered by a frequent value form a dynamic partition for which one basic query is generated. The remaining entities are further partitioned according to frequent values as long as the minimal support per value is achieved. Depending on the actual attribute values the input entities may thus be divided into several partitions of variable size.

- An **attribute-predicate mapping** is a mapping of selected input attributes to their corresponding search engine predicates. Different attributes may map to the same predicate (e.g., the free search predicate) and, in principle, an attribute may map to different predicates. Since very specific queries may lead to a reduced recall the attribute-predicate mapping might only contain a subset of the identified schema mapping correspondences. The attribute-predicate mapping is usually determined beforehand, e.g., based on a manually or automatically determined schema matching [RB01].

- The **search value generation** determines how the predicate search values are derived from the attribute values from the input entities of the same partition. The result is a basic search query per partition. To that end different functions can be applied on the attributes, e.g., to generate phrases (putting a string in quotation marks) or to determine keywords, e.g., by removing stop words from a string. Further transformation functions may be specific to a search engine and we introduce some of them in the example below and in the evaluation section.

- The final **aggregation** is an optional step to combine several basic search queries into one query. The reduced number of queries increases the number of processed input entities per query and may improve search efficiency. For query generators we usually apply a disjunction (OR) of basic queries but our model also supports other types of combination (e.g., AND). The aggregation step is optional also because not all ESEs support such a combination of several queries.

*Example*: For illustration we consider the three fictional publication entities of Table 1 that should be found at the ESE Google Scholar (Scholar). As indicated in Table 1, Scholar provides the predicates author and intitle that match to the attributes authors and title, respectively. A first query generator may use a naïve partitioning and generate a basic query for every publication. The attribute-predicate mapping may only use the title attribute and the transformation function may extract all relevant keywords from the title by skipping stop words. Then the resulting queries without aggregation are $q_1$=intitle(question 42), $q_2$=intitle(don't panic), and $q_3$=intitle(hitchhiker's guide galaxy). If the search engine supports the combination of basic queries with the OR operator, a modified query generator may only generate one query $q=q_1 \vee q_2 \vee q_3$.

A second query generator may use a frequent-value partitioning using the author attribute. Since the entities $s_1$ and $s_2$ share a common author (Smith) they are merged into one partition. The remaining entity $s_3$ forms the second partition. Furthermore, the generator utilizes the predicate author and the transformation function extracts the most frequent name. The resulting queries are therefore $q_1$=author(smith) and $q_2$=author(taylor).

## 4. EVALUATING QUERY GENERATORS

Query generators are a powerful concept to determine an effective and efficient set of search engine queries to find information for a given set of entities. Unfortunately, finding the best query generator(s) is still challenging due to the availability of many query generators per ESE and different performance and effectiveness behavior for different input data sets. By evaluating the query generators on different input sets we obtain insights about their effectiveness and efficiency in dependence of the characteristics of those input sets. This information is then useful for choosing the most promising query generators for on-demand data integration within mashup applications. The selection of query generators may initially be a manual decision by the mashup developer but should eventually become an automatic decision by a mashup infrastructure.

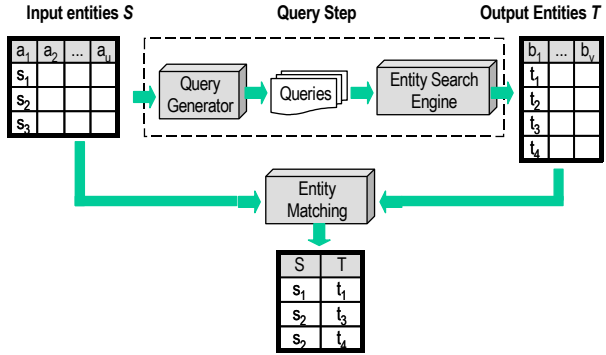

**Figure 1. Search query processing for evaluating query generators**

We propose a general framework for evaluating query generators for entity search engines. The overall approach is illustrated in Figure 1. The query generator to be evaluated is applied on the input set $S$. Each query step deals with the execution of a generated query by the respective ESE $E$. The result is a set $T$ of entities of the same type as the input entities $S$. The entity sets $S$ and $T$ are then matched, i.e., corresponding entities are identified. The match result is a so-called mapping $M \subseteq S \times T$ containing all pairs $(s, t)$ where $s$ and $t$ represent the same real-world entity. Entity matching can be done automatically based on different approaches (see, e.g., [EIV07] for a survey), e.g., the similarity of selected attributes.

The key evaluation idea is that the match result $M$ can be used to automatically derive the quality of the underlying query generator. The match result reveals for what input entities a corresponding entity has been found by the generated queries. Moreover, the match result also identifies irrelevant query results, i.e., entities that do not match any of the input entities. To that end we define the effectiveness measures *Coverage* and *Recall* as follows:

- *Coverage* = |domain($M$)| / |$S$|

- *Recall* = |range($M$)| / | $T_{rel}(S)$ |

*Coverage* is the fraction of $S$ for which at least one matching counterpart is found in the match result, which is denoted as domain($M$) = $\{s \in S \mid \exists t \in T: (s,t) \in M\}$. For example, a *Coverage* of 1 is achieved if for all input entities $s$ at least one relevant entity $t$ appears in the query result. Analogously, range($M$) denotes all $t \in T$ that appear in the match result $M$, i.e., all $t$ having at least one matching counterpart $s$. The *Recall* gives information about to fraction of the found variations of the input entities in relation to all relevant entries $T_{rel}(S)$ that can be retrieved by the search engine for the given input set $S$ (relevant entries include duplicates and spelling variations). However, $T_{rel}(S)$ is usually not given and very expensive to determine (in principle one must execute all possible queries that are somehow related to entities of $S$).

The precision of a query generator can be calculated as follows:

- *Precision* = |range($M$)| / |$T$|

Based on our experience *Precision* is typically less important for query generators than *Coverage* and *Recall*. This is to say that search queries should primarily aim at retrieving all relevant entities (with few queries) even if many irrelevant entities are also obtained. This is because the irrelevant entities can be rather easily filtered away by a subsequent matching step thereby improving precision after the query step. However, the *Precision* might be useful to determine the number of entities that should be requested with one query (see the "next link" evaluation in Section 5).

Measures for characterizing the effectiveness of query generators are only one side of the coin. On-demand data integration also requires fast query response times and, thus, a small number of queries and query requests to retrieve the relevant results. Therefore, a useful measure to determine efficiency is the number of *query requests* a query generator uses for a given input data set.

The number of query requests may be higher than the number of queries if a query returns more entities than can be retrieved within one search engine interaction. Typically, ESEs return a maximum of $z$ (e.g., 10 or 100) entities per query. Obtaining the remaining result entities for the query requires additional query requests, that is, repeatedly follow the "next link". For example, Google Scholar provides at most 100 publications per result page. An author query returning, say, 519 publications would require 6 requests to obtain all result entities.

Since the total number of query requests depends on the number of entities to be found we use the following measure to determine the efficiency of a query generator:

- *Efficiency* = |domain($M$)| / #Requests

    = *Coverage* · |$S$| / #Requests

The definition considers the size of the input set and the number of requests sent to the search engine as well as the coverage. The efficiency measure has an intuitive meaning: it indicates the average number of input entities that are covered per query request. A high value indicates a good efficiency (low number of queries needed to retrieve relevant entities). The combined consideration of coverage is needed to focus on the number of relevant queries (a small number of queries is useless when these queries do not return the requested entities).

## 5. EVALUATION

For our sample evaluation we have chosen Google Scholar[1], a popular ESE for research publications, e.g., conference and journal papers. As indicated in Table 1, it provides a free text predicate (free) and search predicates for title (intitle), authors (author) and year (year). Scholar covers millions of publications but has limited data quality due to an automatic extraction of bibliographic metadata from the reference lists of full text documents (misspelled author names, wrong publication year, etc.). Furthermore, there are frequent duplicate publication entries. Heterogeneous conference and journal names make it difficult to determine all Scholar entries for a particular venue and year. Thus, Scholar is a challenging ESE for evaluating query generators.

As input entities we use subsets of the DBLP[2] data source that indexes more than one million computer science articles. Based on this source we automatically generated 60 test datasets where each dataset consists of either 5, 30, or 100 publications and is assigned to one of the following four categories:

- *Author*: all publications have one common author
- *Title*: all publications have one or more common keywords in the title
- *Venue*: all publications were published at the same venue (conference proceeding or journal volume) in the same year
- *Random*: a random collection of publications

---

[1] http://scholar.google.com

[2] http://dblp.uni-trier.de/

**Table 3. List of evaluated query generators**

| No | Partitioning | Mapping | | Search value gen. | Aggregation |
|---|---|---|---|---|---|
| | | attribute | predicate | | |
| 1 | naïve | title | intitle | keywords | - |
| 2 | naïve | title | intitle | phrases | - |
| 3 | naïve | title | intitle | phrases | OR (2) |
| 4 | naïve | authors<br>title<br>year | author<br>intitle<br>year | gsAuthors<br>keywords<br>value | - |
| 5 | freq. value (author) | authors | author | gsAuthors | - |
| 6 | freq. value (title) | title | intitle | keywords | - |
| 7 | freq. value (2 out of {authors, title, year}) | authors<br>title<br>year | author<br>intitle<br>year | gsAuthors<br>keywords<br>value | - |
| 8 | freq. value (2 out of {authors, title, year}) | authors<br>title<br>year | free<br>free<br>free | gsAuthors<br>keywords<br>value | - |
| 9 | naïve | title | intitle | pattern | - |
| 10 | naïve | title | intitle | pattern | OR (10) |

For every combination of dataset size and category we generated 5 different datasets giving us 3·4·5=60 datasets. We tested ten query generators (see next subsection) on these 60 datasets and saved the matching results and additional information about the query executions and datasets in a data warehouse for evaluation. Due to lack of space we will focus our discussion on the results for datasets of size 30. For the other dataset sizes we observed comparable results. For datasets of size 5 the differences between the query generators were smaller since even the naïve approaches require at most 5 queries.

The matching between the DBLP instances and the retrieved Scholar entities is performed based on a combined similarity value for the three attributes authors, publication title and year. The similarity functions, combination function and match thresholds have been determined with the entity matching tool MOMA [TR07]. All string comparisons are case insensitive. The comparison of author names is based on the last names and the first letter of the first names. To compare two publication years we use the measure $1-(\min(|year_1-year_2|,10)/10)$. Finally, a pair of a DBLP publication and a Scholar publication is added to mapping *M* iff it achieves similarity values of at least 0.5 for the authors, 0.8 for the title, and 1 for the year.

### 5.1 Query Generators

Table 3 shows the 10 query generators used in this evaluation. For each generator all four building blocks (partitioning, mapping, search value generation, and aggregation) are specified.

The first two generators demonstrate the effectiveness of using only one attribute for searching, but with different search values (keywords vs. phrases). They generate exactly one query per input entity (naïve partitioning). Query generator #3 corresponds to #2 but OR-combines two basic queries to cut the number of queries by half. The fourth generator maps three attributes to the corresponding search predicates of Scholar and may therefore be more precise than the first three generators. Query generators #5, #6, #7, and #8 utilize a frequent value strategy, i.e., they pre-analyze the input for a value-based partitioning. Generator #5 groups the input set by the most frequently common authors whereas generator #6 partitions the input publications based on common title keywords. Generators #7 and #8 operate on multiple attributes and build the partitions by identifying two common items (authors, title keywords and/or year) for each partition. Query generator #8 is similar to #7 but uses the free search predicate for all search values.

The generators #9 and #10 utilize a special wildcard feature of Scholar. It can be used to create queries with relaxed phrases (pattern) while still aiming at high query precision. For example, when searching for publication [TR07] the pattern intitle:"MOMA * * * object" can be used instead of the complete title. The pattern is determined by replacing common words by a wildcard character (asterisk) so that the remaining word list still characterizes the publication title unambiguously within DBLP. Since this approach preserves the word order in the title, the results of this strategy are typically more precise than naïve strategies based on keywords.

The selected query generators can illustrate opportunities of our framework but do obviously cover only a subset of all possible approaches. We plan more comprehensive evaluations in our future work.

### 5.2 Evaluation Results

An evaluation of query generators can be accomplished on very different dimensions concerning the characteristics of the test datasets or properties of the query generators. In the following we will first focus on the effectiveness of the query generators and then discuss efficiency which takes the number of query requests into account. The main goal is to identify the best query generators for the four different categories of input data.

Figure 2 illustrates the coverage of each query generator for the four dataset categories. We observe that the best results (up to 0.8) are achieved for generators #1, #2, #3, and #9 which only utilize the intitle search predicate based on publication title[3]. They perform similarly well for all four categories of input data. These good results are a consequence of the fact that the generators mostly use one query per publication, i.e. they require great search efforts. The use of relaxed search values results in a slight decrease of coverage, e.g. both the pattern-based (#9) and the keywords-based (#1) query generators perform slightly more effective than the use of phrases (#2, #3). On the other hand, generator #4 created very restrictive queries so that it misses many

---
[3] Coverage and recall values of 1 cannot be expected since not every DBLP publication is represented in Scholar. We noticed that especially older papers (e.g., publication date < 1995) are mostly not available in Scholar.

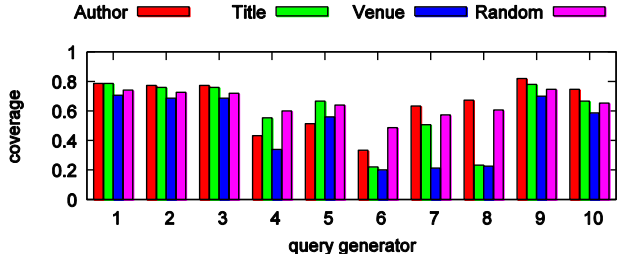

**Figure 2. Average coverage of the 10 query generators per category**

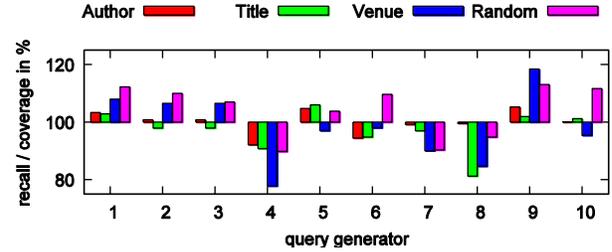

**Figure 3. Comparison of coverage and recall for query generators 1-10 (coverage is normalized to 100%)**

relevant entities resulting in a low coverage (which cannot be compensated by its good precision).

The comparison of query generators #2 and #3 shows that the combined processing of two intitle queries does not lead to a significant quality loss. Generator #10 even reduces the number of queries by a factor 10, while retaining about 87% of the coverage of generator #9. Hence combining several queries has comparatively little impact on effectiveness but does significantly improve efficiency as we will see later in this section. On the other hand, the query generators based on frequent values may also reduce the number of queries (see below) but achieve only a medium coverage. Hence these query generators alone may not be effective enough and may have to be combined with other query generators for sufficient coverage.

In Section 4 we have differentiated between coverage and recall to quantify the completeness of query results. Figure 3 compares these measures by building their ratio. In general, both measures are in a comparable range and are thus useful to evaluate effectiveness. For generators #1, #2, #3, and #9, which performed best for coverage, recall is even higher than coverage (ratio>1). This is because these generators provide the most duplicate entities for the publications in the input datasets. These duplicate entities may contain useful complementing information, e.g., to find all citations of a publication. In particular, the pattern-based query generator #9 benefits from its precise but still approximate queries to find variations of the same publications. By contrast, the recall values for generators #4, #7, and #8 are worse than their coverage. These generators (see Table 3) use three attributes for their queries and, thus, produce very specific queries, hence missing duplicate entities with slightly different attribute values (e.g., due to a typo in the title).

In order to identify the best query generators we also have to take their efficiency into account. For this purpose we introduced a relative efficiency measure which indicates the average number of relevant entities found per query request. Figure 5 shows how the different query generators perform w.r.t efficiency measure. We observe that the naïve query generators without query combination mostly achieve quality values of about 1, since they issue a query per entity (query generator #4 performs worst because of its limited coverage). Better quality values are achieved by combining several basic queries (query generator #3, #10) for all four categories of input data. Query generator #10 combining 10 pattern-based search queries is especially successful by achieving 5-6 relevant entities per query request for all four types of input data.

The query generators based on frequent values are also able to reduce the number of queries and thus to improve efficiency but their quality differs substantially for different categories of input data. Query generator #5 is by far the most efficient generator in category Author, since it needs only one basic author query for all input entities (several query requests are necessary to follow the "next links"). However, this query generator is not efficient for the three other types of input data since they typically have no frequently occurring authors. Query generators #6 and #7 utilize frequent title keywords and are able to improve the efficiency for the second input category (publications with titles sharing the same term). However their efficiency is much lower than for generator #5 on frequent author datasets. This is because author names are relatively distinct and authors typically have a smaller number of publications compared to the number of publications with a given keyword. Input category "venue" is not well supported by the considered query generators on Scholar so that such input entities could be treated similarly than a random set of publications.

Figure 5 also distinguishes the efficiency between the $1^{st}$ request and the average efficiency for all requests. We observe that for the most efficient generators the first request is especially efficient while the remaining requests (e.g., "next link" requests) are still useful but reduce the average efficiency. The high efficiency of the first request is influenced by an apparently good ranking of the considered search engine and is useful if one has to strongly limit the number of search queries. Generators with naïve partitioning, e.g., #1-#3 produce one query per input entity and, thus, only return few query results. In most cases the query execution therefore needs only one request. Hence, the efficiency of the first request corresponds to the average request efficiency.

Finally, we want to explore the usefulness of "next link" query requests. In our evaluation 22% of all requested result pages have offered a next link. The question then becomes whether spending another query request to follow this link will likely return relevant results. To answer this question we have analyzed the precision of the current result page (percentage of relevant result entities) and compared it with the precision of the next result page. This is illustrated in Figure 4 where the x-axis refers to the precision of the current result page and the y-axis indicates the precision of the next result page. Figure 5 also covers the experiments with 100 input entities for which following the "next link" is more relevant than for smaller inputs. Note that for our settings (result page size = 100) there was no result page with more than 60% precision (i.e., more than 60 relevant publications) offering a next link. We observe that the precision of the next result page usually decreases but that in many cases the next page still provides many relevant entities (up to 30% precision). Based on the results we can derive a simple precision criterion to decide whether we should follow the next link. For example, if we want to achieve at least 5% precision in a next result page we should not follow the next link if the precision of the current result page is lower than 15%.

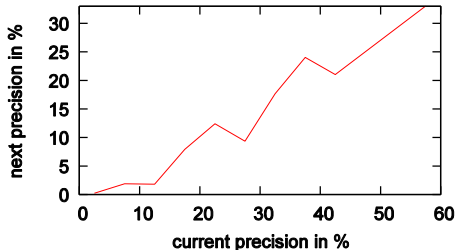

**Figure 4. Precision of "next link" requests**

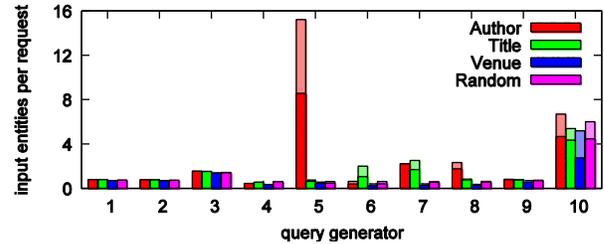

**Figure 5. Efficiency of query generators per category
(complete bar = efficiency for 1<sup>st</sup> request;
lower bar = average efficiency for all requests)**

The given example evaluation illustrates the spectrum of evaluation types that can be realized with our approach to identify the most promising query generators based on their coverage, recall, and efficiency characteristics. In our evaluation scenario query generators using one search predicate usually outperform generators with three predicates. The utilization of specific search engine features, such as pattern-based queries and the combination of several queries, proved to be highly efficient. Furthermore, we observed that next link queries are effective as long as the current result page provides a certain precision level. The use of frequent value query generators is promising depending on the type of input data which would have to be analyzed beforehand. Our results also indicate that a single query generator may not always be sufficient to achieve both good coverage and good efficiency. Hence, there is a need to study the combined use of several query generators or the construction of more sophisticated query generators. For example, one could first use a highly efficient query generator (e.g., number #10 or #5 for author-based publication sets) to quickly present search results to the mashup user and use additional query generators in the background to continuously improve coverage during mashup execution.

## 6. SUMMARY AND OUTLOOK

We presented a flexible query generator approach for querying entity search engines based on a given set of input entities. The provision of query generators facilitates the development of powerful mashup applications requiring efficient access to ESEs. We proposed a generic model of query generators comprising several building blocks for a flexible definition of a search strategy. In addition, we illustrated how query generators can be evaluated in a fully automatic way based on automatic entity matching. We finally presented results of an initial evaluation for a selected search engine to demonstrate the usefulness of our approach. The evaluation approach is useful for developers to identify the most promising generators in terms of efficiency and effectiveness.

In future work we will extend our study of query generators to additional domains and search engines. Furthermore, we will develop adaptive search strategies that make use of multiple query generators.